\documentclass[iop,apj]{emulateapj}

\newcommand{\noprint}[1]{}

\newcommand{\myemail}{buenzli@mpia.de}

\slugcomment{Draft version \today}

\shorttitle{Cloud Structure of the Nearest Brown Dwarfs II}
\shortauthors{Buenzli et al.}

\begin{document}

\title{Cloud Structure of the Nearest Brown Dwarfs II: High-amplitude variability \\ for Luhman 16 A and B in and out of the 0.99 micron F\MakeLowercase{e}H feature}

\author{Esther Buenzli$^{1}$, Mark. S. Marley$^2$, D\'aniel Apai$^{3,4}$, Didier Saumon$^{5}$, \\ Beth A. Biller$^6$, Ian J.M. Crossfield$^4$, Jacqueline Radigan$^7$} 

\affil{$^1$Max Planck Institute for Astronomy, K\"onigstuhl 17, D-69117 Heidelberg, Germany, \myemail}
\affil{$^2$NASA Ames Research Center, MS-245-3, Moffett Field, CA 94035, USA}
\affil{$^3$Department of Astronomy, University of Arizona, 933 N. Cherry Avenue, Tucson, AZ 85721, USA}
\affil{$^4$Department of Planetary Sciences, University of Arizona, 1629 E. University Blvd, Tucson AZ 85721, USA}
\affil{$^5$Los Alamos National Laboratory, Mail Stop F663, Los Alamos, NM 87545, USA}
\affil{$^6$Institute for Astronomy, University of Edinburgh, Blackford Hill, Edinburgh EH9 3HJ, UK}
\affil{$^7$Space Telescope Science Institute, 3700 San Martin Drive, Baltimore, MD 21218, USA}

\begin{abstract}
The re-emergence of the 0.99~$\mu$m FeH~feature in brown dwarfs of early- to mid-T spectral type has been suggested as evidence for cloud disruption where flux from deep, hot regions below the Fe cloud deck can emerge. The same mechanism could account for color changes at the L/T~transition and photometric variability. We present the first observations of spectroscopic variability of brown dwarfs covering the 0.99~$\mu$m FeH feature. We observed the spatially resolved very nearby brown dwarf binary WISE J104915.57-531906.1 (Luhman~16AB), a late-L and early-T dwarf, with HST/WFC3 in the G102 grism at 0.8-1.15~$\mu$m. We find significant variability at all wavelengths for both brown dwarfs, with peak-to-valley amplitudes of 9.3\% for Luhman~16B and 4.5\% for Luhman~16A. This represents the first unambiguous detection of variability in Luhman~16A. We estimate a rotational period between 4.5 and 5.5~h, very similar to Luhman~16B. Variability in both components complicates the interpretation of spatially unresolved observations. The probability for finding large amplitude variability in any two brown dwarfs is less than 10\%. Our finding may suggest that a common but yet unknown feature of the binary is important for the occurrence of variability. For both objects, the amplitude is nearly constant at all wavelengths except in the deep K~I~feature below 0.84~$\mu$m. No variations are seen across the 0.99~$\mu$m FeH feature. The observations lend strong further support to cloud height variations rather than holes in the silicate clouds, but cannot fully rule out holes in the iron clouds. We re-evaluate the diagnostic potential of the FeH~feature as a tracer of cloud patchiness. 
\end{abstract}


\section{Introduction}

At the transition from L to T spectral types ($T_{eff}\approx1200-1300$~K), the clouds in the atmospheres of substellar objects disappear from the photosphere, resulting in drastic changes in the spectra over a small range of effective temperatures of only 100-200~K \citep[e.g.][]{kirkpatrick05}. The process by which the clouds disappear is not yet well understood. A view in which increasing particle sizes lead to rain out \citep[e.g.][]{tsuji03,knapp04}, and thus physically and optically thinner clouds that eventually disappear completely, can broadly explain the color-magnitude evolution \citep{saumon08} and spectral series \citep{cushing08, stephens09} from late L to mid T dwarfs, as well as the luminosities of an L/T transition binary with known dynamical masses \citep{dupuy15}. 

Another mechanism that has been proposed is the break up of cloud layers \citep{ackerman01}. Models of partly cloudy atmospheres \citep{burgasser02,marley10} can also broadly explain the color evolution in the transition, in particular the J band brightening. Additional evidence pointing to cloud disruption was provided by the discovery of high-amplitude variable early T dwarfs \citep{artigau09, radigan12, gillon13}. Furthermore, the strengthening of the FeH feature at 0.99~$\mu$m for early to mid-T dwarfs has been interpreted as a sign of cloud holes \citep{burgasser02}. Iron condenses into clouds and depletes the atmosphere of iron-bearing gases such as FeH. The feature weakens for late type L dwarfs, and as the clouds sink below the photosphere the detectable FeH is expected to decrease further. However, its re-emergence for early to mid-T dwarfs suggests that the cloud deck does not sink gradually, but opens holes into hotter regions in which FeH is not depleted. 

The re-emergence of FeH is based on a fairly small sample of brown dwarfs. Recently, observations of the L/T transition binary 
WISE J104915.57$-$531906.1 (hereafter Luhman~16AB) by \citet{faherty14} showed that the strength of the FeH feature was equal for the L7.5 type component and for the T0.5 type component, even though the T0.5 component is highly variable \citep{gillon13,biller13,burgasser14}, suggesting patchy clouds. Furthermore, spectroscopic variability measurements of Luhman 16B \citep{buenzli15} and other early T dwarfs \citep{apai13} at 1.1 to 1.7~$\mu$m have shown that a mixture of thin and thick clouds, rather than cloud holes into significantly hotter regions, are required to explain the wavelength dependence of the variability amplitude. 

Here we present the first observations of spectroscopic variability of brown dwarfs between 0.8 and 1.1~$\mu$m, covering the 0.99~$\mu$m FeH feature. We observed the Luhman 16AB binary with the Hubble Space Telescope (HST). Luhman 16 is an ideal target due to its very close proximity to the Earth \citep[2 pc,][]{luhman13, boffin14, sahlmann15}, allowing high signal-to-noise observations, and the possibility for resolved observations of both a late L and early T dwarf. Furthermore, it has already been widely characterized \citep{luhman13, gillon13, burgasser13, kniazev13, biller13, boffin14, burgasser14, crossfield14, faherty14, buenzli15, lodieu15, sahlmann15}. We detect significant variability in {\em both} brown dwarfs. We discuss the wavelength dependence of the variability and compare to heterogeneous cloud models, with a special focus on the FeH feature. 

\section{Observations and data reduction}

\begin{figure}
\epsscale{1.15}
\plotone{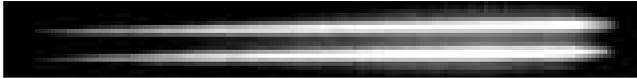}
\caption{A raw first order spectrum on the detector for Luhman 16A (top) and B (bottom), spanning about 0.8 to 1.15 $\mu$m. The dispersion direction is approximately toward north east. The gray scale is logarithmic. 
\label{fig:spec}}
\end{figure}

Observations of Luhman 16A and B were carried out during 5 consecutive HST orbits between 17:20 UT on 2014 November 22 and 00:28 UT on 2014 November 23 (Program 13640). During each 96~min orbit, the target was observable for 50~min with gaps of 46~min when the target was behind the Earth, except for the first gap which was only 30~min. The shorter first gap is a result of scheduling near the continuous viewing zone of HST. We used WFC3 in its infrared channel, obtaining spectral time series of the binary with the G102 grism. The G102 grism spans wavelengths between about 0.8 and 1.15~$\mu$m. We used the 256$\times$256 subarray mode which allowed storage of all observations in an orbit in the WFC3 buffer, optimizing the observing efficiency because no buffer dump was required during the visibility periods. With a pixel size of $\approx$0\farcs13, the field of view was $\approx$30\arcsec x 30\arcsec. The first order spectra, spanning $\approx$155 pixels with a dispersion of 2.36 to 2.51~nm pixel$^{-1}$, were fully captured on the subarray (Fig. \ref{fig:spec}). Several faint spectra of background stars are also visible in the field, but none of them overlap with any of the spectra of our binary target. The grism dispersion direction, given by the orientation of the spacecraft, was set very close to perpendicular to the line connecting the binary, minimizing spectral overlap between the two objects. For wavelength calibration, a direct image in the F098M filter was obtained at the beginning of each orbit to measure the location of the sources. All observations were conducted in staring mode without dithering to avoid errors from pixel-to-pixel sensitivity variations. For the spectroscopic time series, we used the SPARS25 readout mode, obtaining exposures with three non-destructive reads at 0~s, 0.278~s and 22.62~s (NSAMP=2). We took 72 exposures in the first orbit and 75 in subsequent orbits. The cadence, i.e. the exposure time plus overhead for one exposure, was 41~s. We averaged 3 spectra, thus binning to a cadence of 123~s to improve the signal-to-noise ratio (SNR). In total, the time series spanned 7.1~h. Direct images were taken in the SPARS10 readout mode with a read at 0~s and at 0.278~s (NSAMP=1), two such exposures were obtained in each orbit. The maximum number of counts recorded in a pixel in the spectra was $\approx$20,000, well below the regime where image persistence can become significant (above $\approx40,000$ counts). 

Data reduction was carried out in the same manner as in \citet{buenzli12, apai13, buenzli14,buenzli15} and readers are referred to these papers for details. In short, we used a combination of the WFC3 pipeline, custom IDL and Python routines and the PyRAF software package aXe\footnote{http://axe-info.stsci.edu/} which was specifically developed to extract and calibrate slitless spectroscopic data from HST. The zero-read and dark current was subtracted and the images were corrected for non-linearity and gain, as well as bad pixels and cosmic rays. The \texttt{axeprep} routine was used for background subtraction by scaling a master sky frame, and the \texttt{axecore} routine for flat-fielding, wavelength calibration, extraction of the two-dimensional spectra and flux calibration with the G102 sensitivity curve. The extraction width for the spectra was set to 6 pixels. We applied aperture correction based on the values of \citet{kuntschner11}, interpolating between the listed values. Because the beginning of the first orbit is impacted by a well-known ramp effect, we remove the first 20~min, as they are not crucial to our science results. We note that the G141 observations of Luhman 16 in \citet{buenzli15} had to be taken in SPARS10 rather than SPARS25 readout mode because a shorter exposure time was required. With the SPARS10 mode we had found a much larger decreasing ramp effect spanning the first two orbits. Those systematics had been corrected using the non-variability of Luhman 16A in those observations. Such a procedure was not necessary for the new SPARS25 observations.

The separation of the binary has further decreased within the past 12 months, from 1.24\arcsec to approximately 0.9\arcsec. We estimate the flux overlap in the extraction apertures of Luhman 16A and B from the aperture correction values. In y-direction, the two spectra are separated by $6.58\pm0.02$ pixels. With a 6 pixel extraction width, the flux from A present between 3.58 and 9.58 pixels distance from the peak on one side is in the aperture of B, and vice versa. We find that depending on wavelength 2.5-3\% of the flux of the other brown dwarf is present in the aperture, compared to about 90\% of the flux of the brown dwarf on which the aperture is centered. Considering that B is about 40\% brighter in the Y band peak than A and that the maximum variability is less than 10\% for B, less than $0.5\%$ of the measured variability of A can be attributed to originate from B. On the other hand, only $\approx$0.1\% of the measured variability of B can originate from A. 

\section{Results}

\begin{figure}
\epsscale{1.18}
\plotone{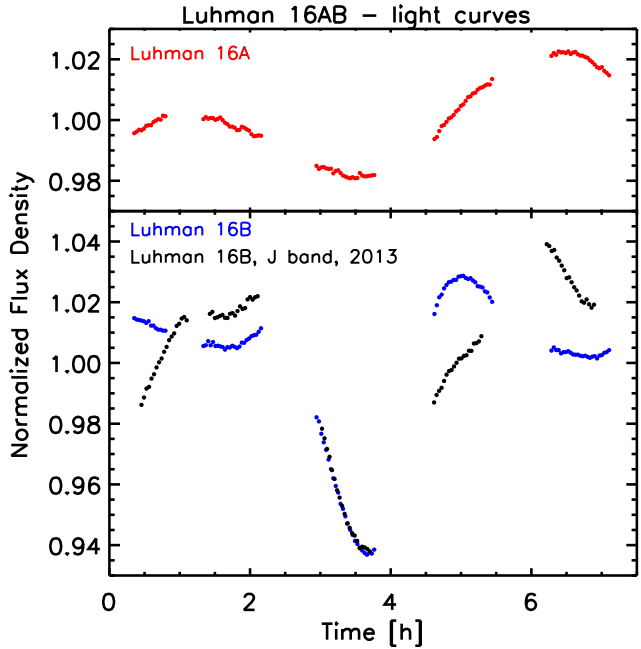}
\epsscale{1}
\caption{Light curve for Luhman\,16 A (top, red) and Luhman\,16 B (bottom, blue) obtained by integrating counts over the fu©ll spectral range of the G102 grism. For comparison, the J band light curve (1.22-1.32 $\mu$m) of Luhman\,16 B from \citet{buenzli15} is also plotted (black), shifted in time to match the phase of the dip. \label{fig:luh16ab_lc}}
\end{figure}

\begin{figure}
\epsscale{1.18}
\plotone{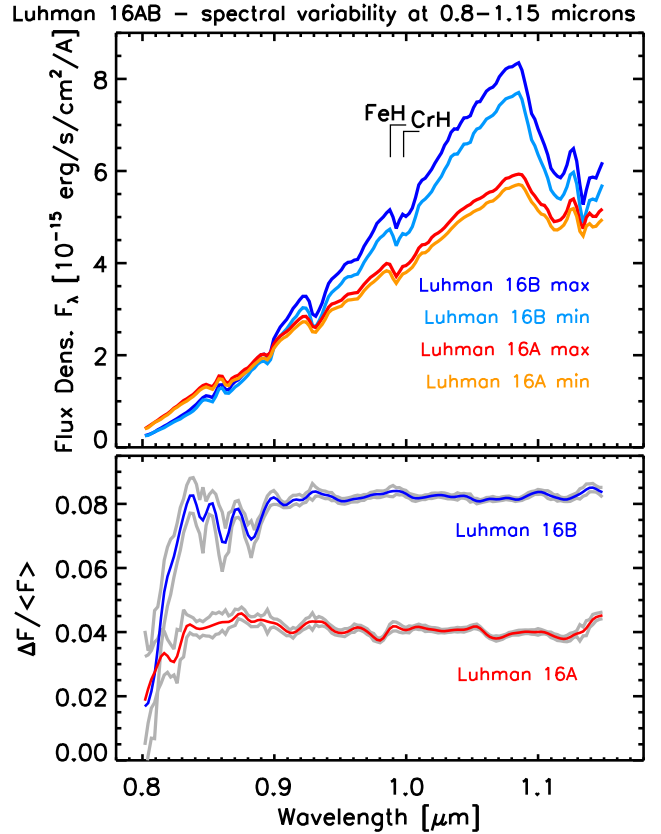}
\epsscale{1}
\caption{HST Observations of Luhman\,16 A and B with the G102 grism. Top: Maximum and minimum spectra averaged over $\approx$15 min. The location of the FeH and CrH features are marked. Bottom: Relative amplitude as a function of wavelength, i.e. the difference of the maximum to the minimum divided by the mean. The $1\sigma$ error is shown as a gray lines. \label{fig:luh16ab}}
\end{figure}

We detect significant variability at all wavelengths for both sources. The integrated light curves are shown in Figure \ref{fig:luh16ab_lc}. The maximum and minimum spectra and the amplitudes as a function of wavelength are shown in Figure \ref{fig:luh16ab}.These observations represent the first unambiguous variability detection for Luhman 16A, although tentative evidence for variability at these wavelengths was previously reported in \citet{biller13}. 

\subsection{Light curves}

Because the variability amplitudes show very little wavelength dependence over the whole measured spectrum, we only discuss light curves integrated over the whole covered spectral region (Fig. \ref{fig:luh16ab_lc}). The peak-to-valley amplitudes are 9.3\% for Luhman 16B and 4.5\% for Luhman 16A. Since we observe a turning point at the maximum and minimum flux values, it is very likely that the maximum and minimum are indeed covered for both objects and we measure the full amplitude. A rapid and deep dip in the flux is observed for Luhman 16B in the third orbit. This dip is very similar to the one seen in the longer wavelength observations taken one year earlier \citep{buenzli15}. At other rotational phases, the light curve shape has changed significantly.  

With a coverage of only 7 h and significant gaps, as well as potential light curve evolution on those time scales, we cannot determine reliable periods. We note a potential period of $\approx$5.15 h for Luhman 16B due to the presence of two identically shaped local minima in the second and fifth orbits. This is slightly longer than $4.87\pm0.01$~h measured by  \citet{gillon13}, but consistent with the $5.05\pm0.1$~h period measured by \citet{burgasser14}. For Luhman 16A, \citet{biller13} suggest a period of 3-4 h from their tentative detection of variability. In our light curve, we find two maxima, although the second maximum is brighter by approximately 2\% of the mean flux. Assuming a similar light curve shape but overall brightening, we roughly estimate the rotation period to be between 4.5 and 5.5 h. Periods below 4.5 h can be excluded unless the light curve shape also changes drastically. The period may also be significantly longer than 7 h in case of a double-peaked light curve. However, the $v \sin{i}$ measurement of 17.6 km/s \citep{crossfield14} indicates a maximum rotation period of 7~h assuming a radius of 1~R$_{J}$. Therefore, a double-peaked light curve for Luhman 16A is not very likely. The potentially very similar rotation periods for Luhman~16A and B, together with different projected equatorial rotational velocities ($26.1\pm0.2$ km/s for B and $17.6\pm0.1$ km/s for A) measured by \citep{crossfield14}, would indicate some misalignment between the rotation axes of the binary components. 

\subsection{Spectroscopic variability}

\begin{figure*}
\epsscale{1.1}
\plotone{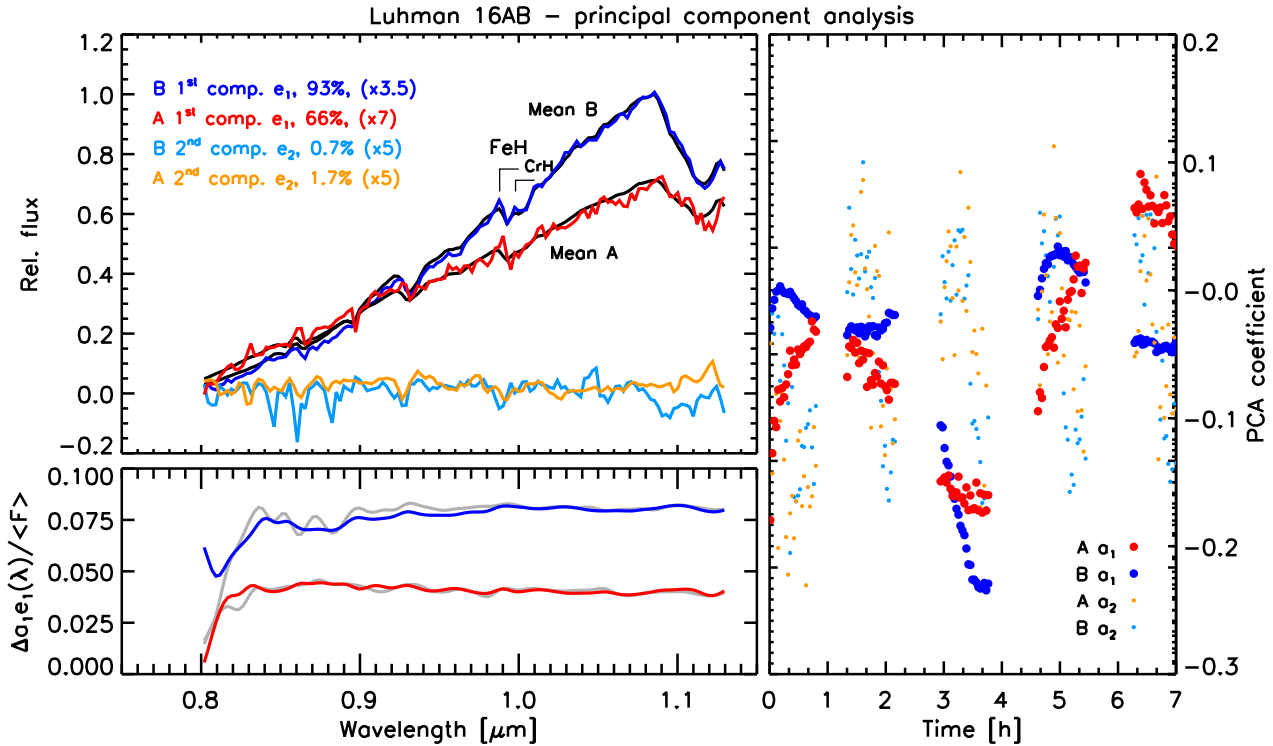}
\caption{Principal component analysis of Luhman\,16 A (red) and B (blue). Top Left: First and second PCA components compared to the mean flux, normalized to the maximum flux of the B component. The first PCA Components are scaled to allow comparison with the mean flux, the second PCA components are scaled for better visibility. Bottom left: Variability amplitude (maximum difference of coefficients multiplied by the component and divided by the mean) derived from the mean and first component compared to the variability amplitude from the maximum and minimum spectrum (gray, lines from Fig.\,\ref{fig:luh16ab}). Right: Coefficients of the first and second components. 
 \label{fig:pcaAB}}
\end{figure*}

The wavelength dependence of the variability amplitude is largely flat for both sources. No significant deviation in the amplitude is seen in the FeH feature at 0.99~$\mu$m or in most other absorption features. Only at wavelengths $<0.83$~$\mu$m the amplitude decreases in both sources. This coincides with the onset of the very deep K I feature centered at 0.77~$\mu$m. However, it is also near the edge of the spectrum where the throughput as well as the flux is low and errors are larger. Additionally, small dips in the amplitude are seen for Luhman 16B between 0.85 and 0.9 $\mu$m. 

We estimate the error on the variability amplitude as a function of wavelength by calculating the standard deviation of the flux in neighboring pixels, which belong to one resolution element, after normalizing with a mean spectrum over the full time series. While the error is below about 0.1\% at $\lambda > 0.9$~$\mu$m for both sources, it increases rapidly toward shorter wavelengths (see Fig. \ref{fig:luh16ab}). At most wavelengths, the SNR of the variability amplitude is between 20 and 50 for Luhman\,16 A, and can rise above 100 for Luhman\,16B. The SNR remains above 5 down to 0.82 $\mu$m for both objects, where the amplitude has decreased by about 1\% for Luhman\,16A and by 3\% for Luhman\,16B. To test whether the decreasing amplitude is not only a result of increasing error due to the lower signal, we fit a linear model between 0.835 and 0.8 $\mu$m and compare the chi-squared value to that for a model where the amplitude remains constant at a fixed value of 4\% or 8\% respectively. The chi-squared value is larger by 25 for Luhman 16A and by 100 for Luhman 16B. Despite the two additional degrees of freedom for the linear model with a slope, we can reject the null hypothesis that the constant model is valid at $>99.999\%$ confidence. The decrease in amplitude as a function of wavelength therefore appears to be highly statistically significant for both objects. This statistical test assumes Gaussian errors. We cannot fully exclude a systematic effect near the grism edge that would introduce correlated errors. However, no similar amplitude change is seen at the other edge of the spectrum.

The achieved precision is comparable to that for the G141 grism \citep{buenzli15}a, except at the shortest wavelengths where the flux drops dramatically. This is mainly due to the fact that Luhman~16A and B are very bright and the G141 observations were less efficient due to very short exposure times. For fainter brown dwarfs, for which the observing efficiency is comparable in both grisms, the G102 observations will typically have lower SNR due to the overall lower count rate. 

We apply a principal component analysis (PCA) to determine the number of variable components, analogous to the analysis in \citet{buenzli15} for the WFC3/G141 observations of Luhman 16B (Fig. \ref{fig:pcaAB}). As was the case at longer wavelengths, a single variable component on top of the mean can explain about 93\% of the observed variability of Luhman 16B. The higher order components can largely be attributed to noise. The flux $F(\lambda,t)$ can therefore approximately described as $F(\lambda,t) \approx \langle F(\lambda,t)\rangle _t + a_1(t) e_1(\lambda)$, where $a_1(t)$ is the first coefficient and $e_1(\lambda)$ the first principal component (eigenvector). 

The first component $e_1(\lambda)$ is essentially proportional to the mean flux, as expected for a flat amplitude as a function of wavelength. The reconstructed time series using only the first component confirms the drop in amplitude at $<0.84$~$\mu$m and the small dip between 0.85 and 0.9 $\mu$m. Additionally, a slight gradient at wavelengths $<1$~$\mu$m is revealed. 

The PCA analysis provides similar results for Luhman 16A, although the first component accounts only for about 66\% of the variability. Nevertheless, all other components account for well below 2\% and are therefore also mostly noise. Again, the wavelength dependence of the amplitude is nearly flat, except for a small dip at about 0.85 $\mu$m and a rapid decrease below 0.83 $\mu$m. There is a very small gradient in the opposite direction than for Luhman 16B, with a $\sim$0.5\% decrease in amplitude from 0.88 to 1.13 $\mu$m. 

\section{Discussion}

\subsection{High amplitude variability for both components of a binary}

The discovery that Luhman~16A is unambiguously variable with an amplitude of about 4\%, at least some of the time, while Luhman~16B is has the second largest J band amplitude of any known brown dwarf, poses the question whether the Luhman 16 system has any particular feature that may favor the occurrence of high amplitude variability over average field dwarfs. 

We first estimate the statistical probability of finding a variable late-L and early-T dwarf assuming independent probabilities. We use the combined sample from the \citet{radigan14a} and \citet{wilson14} surveys, adjusted for the re-analysis from \citet{radigan14b}, to derive variability frequencies. Both were large ground based photometric surveys in J band. Although our variability detection for Luhman~16A is in the Y band, the measured wavelength dependence and models suggest that we would expect a similar amplitude in the J band (cf. \ref{sect:comp} and \ref{sect:models}). From all the sources, we pick the spectral type range L6-L9 to estimate the variability fraction of late-L types like Luhman~16A, and L9.5-T2.5 for early-T types like Luhman~16B. We remove 5 targets with noisy observations where the detection limits might not have allowed a confident detection of the variability of Luhman~16A. There are 6 known binaries from high-resolution imaging in the sample, we count individual components if they fall into our selected spectral type range. None of the binaries show variability. In total, we find 19 late-L and 16 early-T dwarfs to match our criteria. Of those, 2 late-L and 5 early-T are variable at $>1\%$ level with at least $>95\%$ confidence. Following \citet{burgasser03} we calculate the probability distribution for the variability fraction for the late-L and early-T dwarfs using binomial statistics. We then find the combined probability for finding both a variable late-L and early-T dwarf to be $3.3^{+4.3}_{-1.3}$\%. In reality, the probability may be a bit higher because variability amplitudes are known to vary with time and may sometimes result in non-detections even if the object is variable at other times. This is clearly the case for Luhman~16A. 

Nevertheless, at a probability on the order of 10\% or less, finding these large amplitudes in both objects may be more than chance. One possibility is the spin-axis: Luhman~16B is most likely seen near equator-on and any heterogeneities would therefore translate into the maximum possible amplitude. The likelihood for B to be variable may therefore be higher than average. For Luhman~16A the case is less clear, since its possible rotation period may indicate that its axis is somewhat misaligned from B. However, without a reliable period we cannot confidently conclude on the orientation of the spin axis. 

Other shared properties of the binary are the age and initial composition. Furthermore, the objects have relatively similar effective temperatures and for both of them non-equilibrium chemistry appears to be important (cf. \ref{sect:models}). While we currently do not have a large enough sample to look for correlations between these factors and the occurrence of variability, it is possible that some of them contribute to a higher probability of variability. 

\subsection{Comparison to other observations}
\label{sect:comp}

Because of an overlap in spectral coverage between the G141 and G102 observations, we can directly compare the variability amplitude obtained one year apart at least for a small wavelength range. For Luhman 16B, a variability amplitude of about 11\% was found at 1.1-1.15~$\mu$m, slightly larger than in the J and H band peaks. We now find an amplitude of about 9\% at 1.1-1.15 $\mu$m, remaining equal to shorter wavelengths. Assuming that only the overall brightness level has changed, but not the spectral dependence of the variability, we can conclude that the variability amplitude is slightly larger in the Y band peak than in the J and H bands. This is consistent with the results found in \citet{burgasser14}, who found a gradual decrease in the amplitude with wavelength across the near-IR, and the unresolved simultaneous photometry observations by \citet{biller13}. On the other hand, spatially resolved observations from the \citet{biller13} study obtained in a different night showed a large amplitude in H band but no variability in J band, which is very difficult to explain with current models. 

Interestingly, no significant variability $(<0.5\%)$ was found for Luhman 16A over the full G141 wavelength range, including the 1.1-1.15 $\mu$m region. In the new G102 observations however, the variability amplitude is about 4\% in that wavelength region and at shorter wavelengths. This suggests that Luhman 16A goes through periods of quiescence and periods of variability. Extrapolating the measured variability amplitude to longer wavelengths using the spectral dependence of the variable L dwarfs \citep[2M1821, 2M1507][]{yang15} would suggest a variability amplitude of about 2-4 \% in the J and H band. 

The fact that Luhman 16A is variable at a level of a few percent with a likely rotation period that is very similar to that of Luhman 16B, and the fact that its amplitude appears to change with time, may perhaps account for some of the long-term changes in the light curves previously attributed to Luhman 16B by \citet{gillon13}. Furthermore, the interpretation of the light curves is not as straight-forward as if one of the two objects was non-variable. For example, it cannot simply be assumed that the variability amplitude of Luhman~16B was twice that measured in the unresolved observations, as some peaks and valleys in the light curves of Luhman~16B may either be reduced or enhanced by the variability of Luhman~16A. It should also serve as a caution to observational programs that aim to measure the variability of Luhman 16B using Luhman 16A as non-variable reference, as done in \citet{burgasser14}. Although it may have been a valid assumption at that particular point in time, it is not possible to validate this. Nonetheless, with the amplitude of Luhman 16B being at least twice as large as for Luhman 16A, it will remain the dominant contributor to the combined variability. 

Our observations suggest the persistence or re-emergence of a dark spot on Luhman 16B over the time scale of one year. It may be the same prominent dark spot that is also present in the Doppler imaging observations by \citet{crossfield14} in the K band.

Spatially resolved time series of Luhman 16A and B at wavelengths $<1$~$\mu$m have so far only been obtained by \citet{biller13} in the r', i' and z' filters. The z' filter covers approximately the same wavelength range as our G102 observations. \citet{biller13} found the z' band light curves of Luhman 16B to be in phase with the J and H band light curves, this is consistent with our observations, since observations with the G141 did not show phase shifts between the 1.1-1.15 $\mu$m region and the J and H bands. On the other hand, the i' band light curve was anti-correlated with the z-band light curves. Our observations only reach the edge of the i' band, but the steeply decreasing amplitude below 0.85 $\mu$m may point to a phase reversal in the deep potassium feature that dominates the i' band. On the other hand, the tentative detection of i' and z' band variability in Luhman 16A had these light curves in phase, with amplitudes of 3\% in the z' band and 2\% in the i' band. However, we observe a similar drop in the amplitude below 0.85 $\mu$m, from about 4 \% to below 2\%. Considering the very similar variability behavior across the 0.8-1.1 $\mu$m region for Luhman 16A and B, the fact that only Luhman 16B shows a phase reversal in the i' band is somewhat surprising. 

A recent survey by \citet{heinze15} has indicated that some T dwarfs may be much more highly variable in the optical than the near-infrared. They found relatively high amplitude optical variability in two T dwarfs for which lower-amplitude variability had previously been detected in the near-IR. However, because observations were not taken simultaneously, it is also possible that the amplitudes have simply varied in time. Our observations suggest this to be the case for Luhman 16A. The simultaneous monitoring by \citet{biller13} already suggested that Luhman 16B had similar optical and near-IR amplitudes, although an unexplained low amplitude was found in one night only in the J band. Thanks to the overlapping wavelength region in the G141 and G102 grism, our observations confirm that the amplitude between 0.8 and 1.6 $\mu$m remains fairly constant, with the exception of a decrease in the 1.4 $\mu$m water band. Simultaneous optical and near-IR observations of the \citet{heinze15} sample would therefore be crucial to understand whether some objects indeed show strong differences between these wavelengths. If so, it might point to a different origin of the variability than the thickness variations of the silicate clouds we propose in this paper (cf. Sect. \ref{sect:models}) and in \citet{buenzli15}. That might be the case in particular if the variability of these objects originates predominantly from the dark 0.7-0.8 $\mu$m region dominated by the strong potassium opacity, especially if little variability is present at 0.85-0.9~$\mu$m where the flux is much higher. 

\subsection{Comparison to atmospheric models}
\label{sect:models}

\begin{figure*}
\epsscale{1}
\plotone{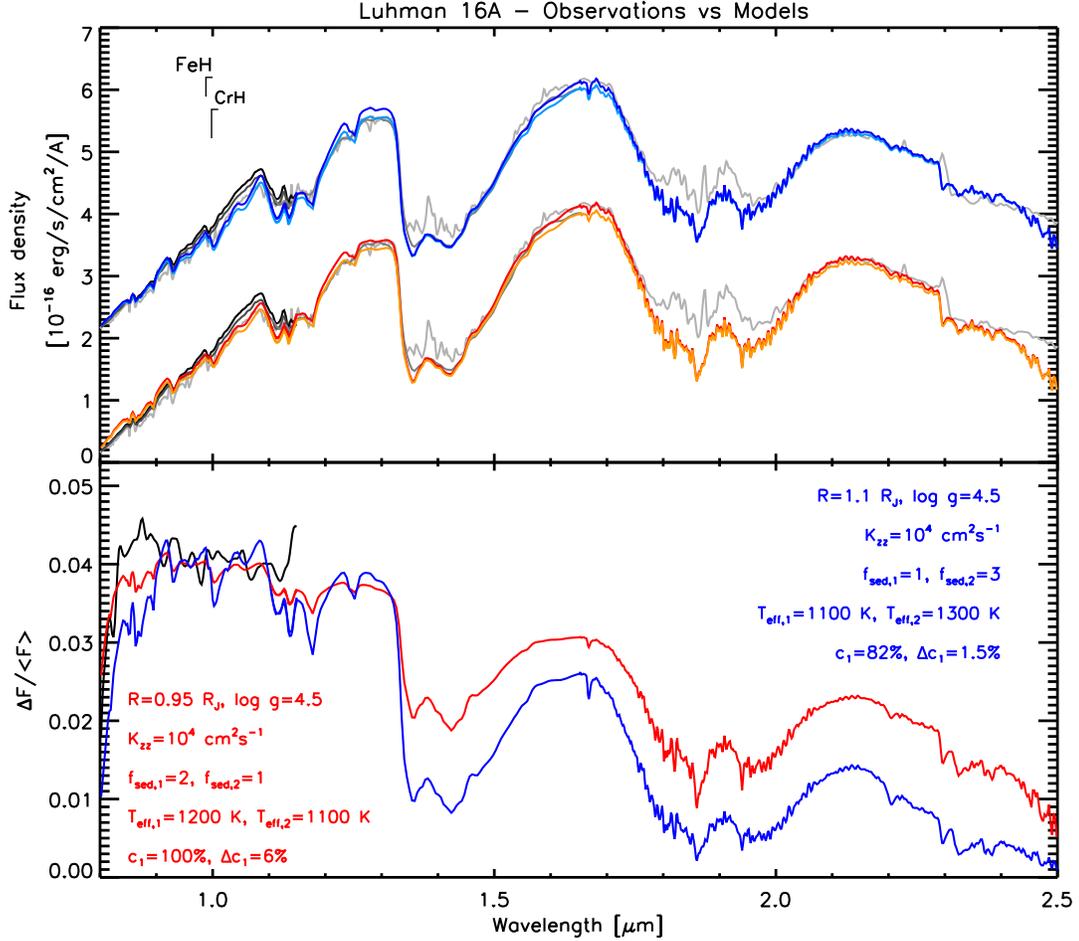}
\caption{Comparison of observations and models for Luhman 16A. The two best 2-component models (blue and red) are shown with corresponding model parameters given in the Figure. Top: The maximum (dark blue and red) and minimum spectrum (light blue and orange), i.e. as given by the different covering fractions on the two hemispheres, for each model compared to observations. For clarity, the observations and models for one of the model cases has a vertical offset of $2\times10^{-16}$ erg/s/cm/\AA. The observations are shown in different shades of gray: The new G102 observations from 0.8 to 1.15 $\mu$m are shown in black (maximum spectrum) and dark gray (minimum spectrum). The G141 \citep[1.1-1.65 $\mu$m,][]{buenzli15} observations are shown in a medium gray and the FIRE spectrum  \citep{burgasser13} in light gray. The ground-based FIRE observations are strongly impacted by telluric lines in the water absorption bands. Bottom: The relative amplitude as a function of wavelength, i.e., the difference of the maximum to the minimum divided by the mean, for the observations (black) compared to the two models (red and blue), and model predictions for longer wavelengths.  \label{fig:model_16A}
}
\end{figure*}

\begin{figure*}
\epsscale{1}
\plotone{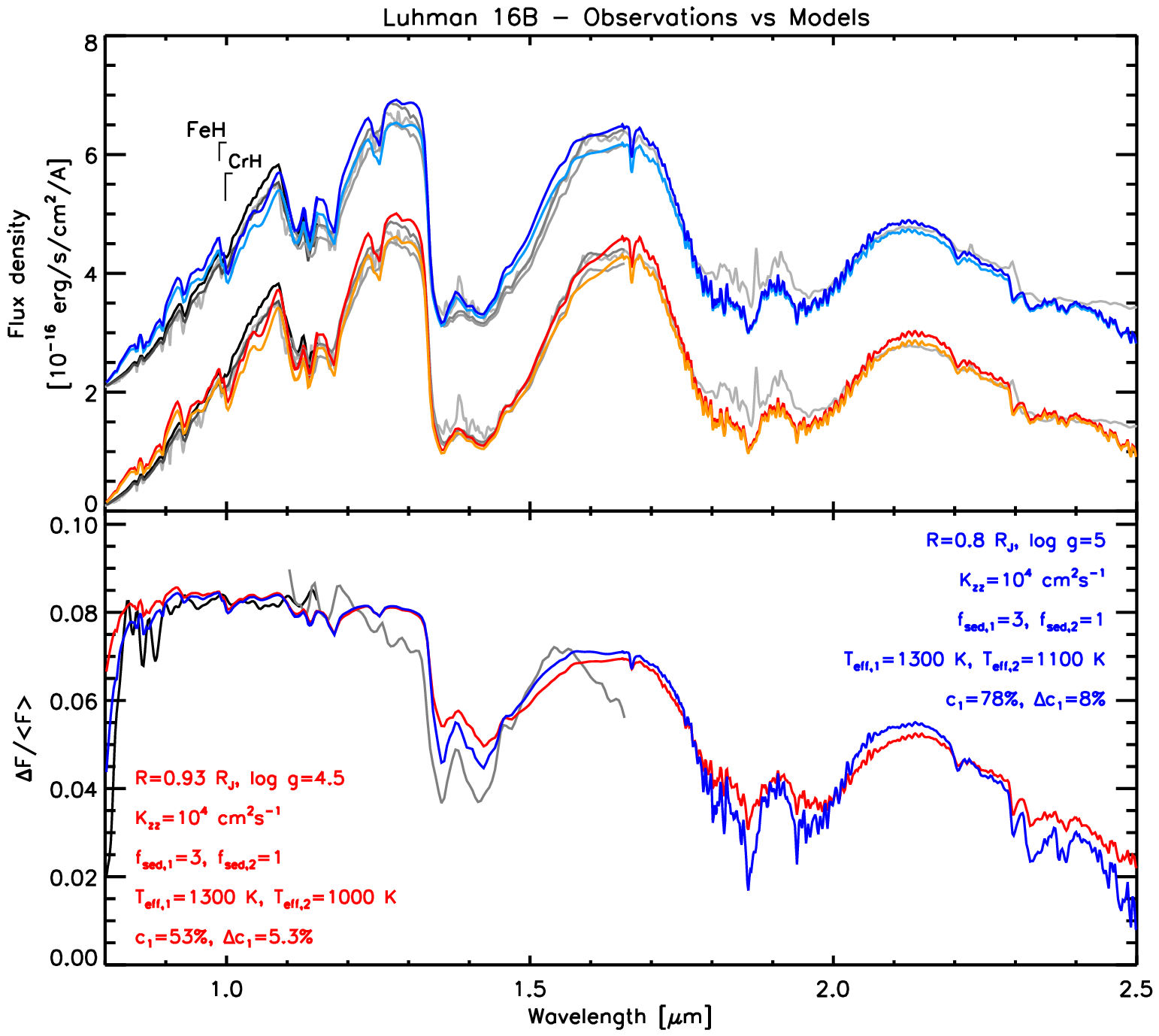}
\caption{Comparison of observations and models for Luhman 16B. Two best 2-component models (blue and red) are shown with corresponding model parameters given in the Figure. Top: The maximum (dark blue and red) and minimum spectrum (light blue and light red), i.e. as given by the different covering fractions on the two hemispheres, for each model compared to observations. For clarity, the observations and models for one of the cases has a vertical offset of $2\times10^{-16}$ erg/s/cm/\AA. The observations are shown in different shades of gray: The new G102 observations from 0.8 to 1.15 $\mu$m are shown in black (maximum spectrum) and dark gray (minimum spectrum). The G141 \citep[1.1-1.65 $\mu$m,][]{buenzli15} observations are shown in a medium gray and the FIRE spectrum  \citep{burgasser13} in light gray. The ground-based FIRE observations are strongly impacted by telluric lines in the water absorption bands. Bottom: The relative amplitude as a function of wavelength, i.e., the difference of the maximum to the minimum divided by the mean, for the observations (black) compared to the two models (red and blue), and model predictions for longer wavelengths.  \label{fig:model_16B}
}
\end{figure*}

We use the same models as in \citet{buenzli15} to determine whether they are able to also explain the observations at 0.8-1.15 $\mu$m. They are based on the \citet{ackerman01} cloud models and have been updated with new opacities, among them FeH \citep{saumon12}. The models and parameter range covered are described in \citet{buenzli15}. In brief, two models are linearly combined. These can have different effective temperatures $T_{\mathrm{eff,1}}$ and $T_{\mathrm{eff,2}}$ and different cloud parameters $f_{\mathrm{sed,1}}$ and $f_{\mathrm{sed,2}}$. A smaller $f_{sed}$ value corresponds to a vertically and optically thicker cloud. 
Gravity $\log{g}$ and the disequilibrium parameter $K_{zz}$ have to be equal for the two models. The weights in the linear combination are set using the covering fraction $c_1$ of the warmer model on the hemisphere where the emitted flux is maximal, and the change in covering fraction $\Delta c_1$. Then, the combined flux $F_{\mathrm{max}} = c_1 F_1 + (1 - c_1) F_2$ and  $F_{\mathrm{min}} = (c_1-\Delta c_1) F_1 + (1 - (c_1-\Delta c_1)) F_2$. 

For Luhman 16A, the non-variability in the G141 observations and the near-IR spectrum had suggested a homogeneous cloud layer with intermediate cloud thickness $f_{sed}=2$ and $T_{eff}=1200$~K. Because Luhman~16A is now found to be clearly variable, we re-fit the models using the full parameter space and allowing for two different surfaces as for Luhman~16B. We find two potential solutions that are a good fit to the average spectrum and show a relatively flat variability amplitude between 0.9 and 1.1 $\mu$m as well as a sharp decrease below 0.85 $\mu$m. These are shown in Figure \ref{fig:model_16A}. Both best models for Luhman~16A have $\log{g} = 4.5$ and $K_{\mathrm{zz}}=10^4$~cm$^2$s$^{-1}$. The best-fit model remains the model where one hemisphere is fully homogeneous with $f_{sed}=2$ and $T_{eff}=1200$~K, while on the other hemisphere 6\% coverage fraction with cooler, thicker clouds ($f_{sed}=1$ and $T_{eff}=1100$~K) is introduced. An alternative possibility is a model where the cool, thick clouds ($f_{sed}=1$ and $T_{eff}=1100$~K) dominate with 82\% coverage fraction, with the remaining 18\% being warmer, thinner clouds ($f_{sed}=3$ and $T_{eff}=1300$~K). In this case, the change in thick cloud coverage fraction from one hemisphere to the other is only 1.5\% to obtain a variability amplitude of about 4\%. This model also results in a larger radius, 1.1 $R_{J}$, as compared to 0.95 $R_{J}$. The larger radius is more in line with the expected evolution radius \citep{saumon08}. However, for this model, an amplitude decrease of about 0.5\% is predicted for the CrH band, and a decrease of 1\% below 0.9~$\mu$m, which is not seen in the observations. We note that even thinner clouds or fully non-cloudy regions would result in even stronger amplitude differences inside and outside of absorption bands, while more thick or cooler clouds would make the spectrum too red. 

Overall, the best-fit models indicate that the near-IR variability of Luhman 16A, when present, would show a similar wavelength dependence as that of Luhman 16B, with a decrease of the variability amplitude in the water band and a slight gradient across the J and H band. However, the two Luhman~16A models could be clearly distinguished if variability could be measured also at 1.1-1.7 $\mu$m and/or in the K band. The variability amplitude is predicted to be significantly lower in the 1.4 $\mu$m water band and K band for the more heterogeneous model with a larger fraction of cool, thick clouds. 

For Luhman 16B, the spectral dependence of the variability in particular across the 1.4 $\mu$m water feature had suggested a mixture of thin, warm ($f_{sed}=3$ and $T_{eff}=1300$~K) and thick, cool ($f_{sed}=1$ and $T_{eff}=1000$~K or 1100~K) clouds, where the coverage fraction was somewhat degenerate with other parameters such as the radius and surface gravity. To fit the new observations and the overall spectrum, we find that we require the same overall model parameters except for different relative coverage fractions (Fig. \ref{fig:model_16B}). The coverage fraction $c_1$ (fraction of the warmer, thinner cloud on the hemisphere with maximum flux) is reduced by 3-7\% to 53 or 78\% coverage with the thinner clouds, and the difference in coverage fraction $\Delta c_1$ between the two hemispheres is reduced by 0.7-2\% to 5.3 or 8\%, depending on the model.  Both models result in a very flat spectral dependence of the amplitude, including across the FeH feature, very similar to what was observed. The similarity of the model results for the two different wavelengths and times suggest that it is primarily the spatial distribution of the thin and thick clouds that changes in time, and not the intrinsic cloud properties of the two components. 

The steep decrease in the amplitude below 0.84 $\mu$m is better represented by the higher surface gravity model, but the model fits cannot conclusively determine the surface gravity of the object. The overall fit of the mean spectrum between 0.8 and 1.15 $\mu$m is not optimal. In particular, the spectral slope is underestimated by the models, and a few absorption features are predicted to be deeper than measured, especially the CrH feature at 1.0 $\mu$m. As is the case for Luhman 16A, introducing fully non-cloudy regions would result in much more variable amplitudes as a function of wavelength. That case had already been shown unfeasible to explain the spectroscopic variability measured in the G141 grism at 1.1-1.7~$\mu$m \citep{buenzli15}.

The modeling results suggest that the cloud structure of Luhman 16A and B may be very similar in terms of cloud thickness and temperatures, as both can be fit by model combinations of $f_{\mathrm{sed}}=1$ and 3 and $T_{\mathrm{eff}}=1100$ and 1300 K. The main difference in that case is that the fractional coverage of thicker, cooler clouds dominates on Luhman~16A, while that of the thinner, warmer clouds dominates on Luhman~16B. It would suggest that at least in the early L/T transition from late-L to early-T dwarf, the spectral changes could be explained by increasing the fractional coverage of thinner clouds. This is similar to the increasing fractional coverage of cloud holes as suggested in \citet{marley10}, except that having remaining thin clouds instead of deep holes explains the spectrally flat variability ampliudes and keeps the FeH strength equal. The periods of non-variability could be explained by times when the coverage fraction of the two cloud types is more homogeneously distributed along both hemispheres, as differences of the order of 1\% are needed to introduce an easily measurable variability amplitude.

\subsection{FeH as a probe of cloud patchiness?}

The FeH molecule contains the cloud-forming element Fe, and the condensation of Fe into a cloud dramatically decreases the abundance of FeH in the gas phase. The observed strengthening of the feature from late-L to mid-T spectral type has been suggested as evidence that cloud holes appear at the L/T transition \citep{burgasser02, lodders06, marley10}. This line of reasoning assumes that holes would allow flux from well below cloud base, where Fe is still present as gaseous FeH, to emerge. Spectra of alternatively clear and cloudy patches would then presumably show variations in the FeH absorption band depth. Alternatively, \citet{cushing08} showed that a sequence of models with increasing $f_{sed}$ value could also reproduce the strengthening of FeH. Therefore, even patchiness with different cloud thickness, i.e. different sedimentation efficiency, could also be expected to result in variability in the FeH band strength. However we do not observe any differences in the variability amplitude within the FeH feature and outside, both in the late-L Luhman~16A and the early-T Luhman~16B. The spectral modeling of the variability for both sources (Sect. \ref{sect:models}) is incompatible with cloud holes, but does suggest patchiness with different cloud thickness. What are the implications of this result on the model of strengthening of the FeH band across the L/T transition based on cloud patchiness?

There are a few complications that can affect the FeH feature strength that we must consider. Even above the cloud base some FeH remains in the gas in equilibrium with the condensate \citep{visscher10}, although FeH is mostly found below the cloud deck. Furthermore other opacity sources, notably gaseous K and CrH, are also important in this spectral region. So the visibility of the FeH band is not entirely modulated by the cloud but rather by the interplay of several factors. This complexity is evidenced by the fact that both Luhman 16~A and B have the same FeH feature strength \citep{faherty14}, despite Luhman~16B being being 3 sub-types later and with a thinner cloud deck than Luhman 16~A. Although the scatter in observed FeH strengths for late-L and early-T dwarfs was too large to suggest definite strengthening already at those spectral types in \citet{burgasser02}, and the sample was small, the co-eval nature of the binary helps to confirm that the bluer color and larger variability for an early-T compared to a late-L at the same age and initial composition do not necessarily coincide with the strengthening of FeH or different variability in the FeH feature.  

\begin{figure}
\epsscale{1.2}
\plotone{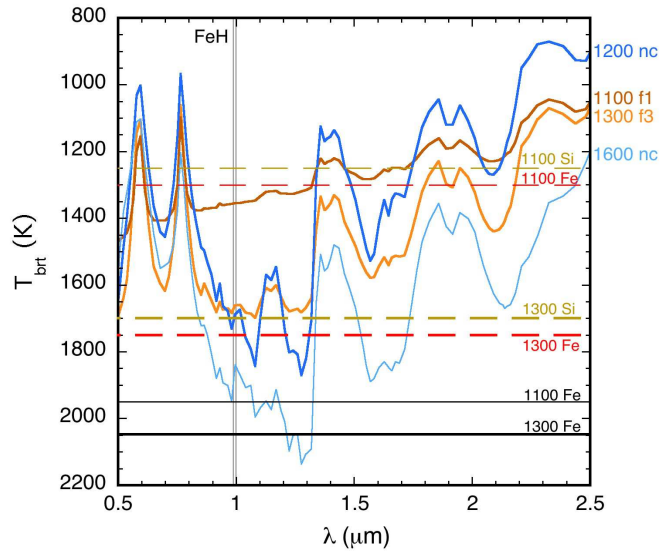}
\caption{Brightness temperature as a function of wavelength for different atmospheric models for Luhman~16B, as labeled along the right hand side. Numbers refer to model effective temperature while `nc' denotes cloudless models; f1 and f3 refer to models with cloud sedimentation efficiency $f_{sed}$=1 and 3, respectively). All models assume $\log g = 5$. The two cloudy models are the same as those that comprise the cloudy case shown in the bottom right of Figure \ref{fig:model_16B}. For the two cloudy models the approximate bases and tops ($\tau \sim 1$) of each model's iron cloud are shown by the horizontal solid and dashed lines, respectively. The top of each model's silicate cloud is also shown as a dashed line, for these models the silicate cloud base is at a similar temperature to the iron cloud base. The FeH band is also indicated by the vertical pair of lines. 
 \label{fig:FeH}}
\end{figure}

To help illustrate the depth from which flux emerges from the atmosphere as a function of wavelength for different scenarios, in Figure \ref{fig:FeH} we present brightness temperature spectra for the two cloud components of one of the Luhman~16B models shown in Figure \ref{fig:model_16B}. We also plot two cloudless models that were not able to fit the observations, but illustrate how the reached depth changes when the cloud opacity is removed. Here brightness temperature serves as a proxy of atmospheric depth from which photons at a given wavelength escape. In reality the contribution function for each wavelength spans a range of pressures around the level in the atmosphere where the local temperature is equal to the brightness temperature. The approximate locations of the model cloud base and cloud tops are shown. Although cloud top is arbitrarily denoted at $\tau\sim 1$, additional cloud opacity is found above these lines. 

Figure \ref{fig:FeH} makes clear that flux from the model which comprises most of the composite flux shown in Figure \ref{fig:model_16B}, $f_{\rm sed}=3$, $T_{\rm eff}=1300\,\rm K$, emerges from above the nominal cloud top and well above the Fe cloud base. For the thick cloud component, $f_{\rm sed}=1$, $T_{\rm eff}=1100\,\rm K$, flux emerges from just below the Fe cloud top, but still far above cloud base. Even for a comparable cloudless model (here with $T_{\rm eff}=1200\,\rm K)$, which highlights the contribution of the gas opacity in the atmosphere, the high molecular opacity at 1~$\mu$m prevents probing to the typical cloud base levels. Only in a much hotter (1600 K) cloudless model would the gas opacity allow emergent flux at 1~$\mu$m to reflect atmospheric conditions below cloud base. This figure leads us to conclude that for objects as cool as Luhman 16~A and B even the complete absence of Fe cloud opacity would not result in an FeH absorption feature from the deep atmosphere to be fully resurgent.

This conclusion, which is contrary to earlier suggestions for using FeH as a tracer of cloud holes or patchiness, stems mostly from a better understanding of the alkali opacities in the same wavelength region and a better atmospheric models incorporating cloud opacity than were available at the time of \citet{burgasser02} and \citet{cushing08}. Certainly for higher effective temperatures, particularly early- to mid-L variable dwarfs, the band strength of the $1\,\rm\mu m$ FeH band may well be variable. 

For mid-T dwarfs, where clouds appear to have mostly cleared out from the photosphere, the situation may be different than at the onset of the L/T transition. There, the strengthening of FeH is more robustly observed by \citet{burgasser02}, even though we are most likely not able to see into potential holes in the Fe clouds. Since the silicate cloud top is predicted by our cloud model to overlie the iron cloud top, the silicate cloud may, counterintuitively, be of greater importance than the Fe cloud to the FeH band depth. As Figure \ref{fig:FeH} makes clear, the tops of the combined silicate and iron clouds greatly influence the depth from which flux near $1\,\rm\mu m$ emerges from the atmosphere. As the cloud opacity in the photosphere drops significantly towards the end of the L/T transition $(f_{\rm sed} \gtrsim 4)$, more of the remaining FeH gas may become visible. Such a gradual sinking of the cloud was initially ruled out by \citet{burgasser02} under the assumption that the FeH was depleted above the cloud. 

Alternatively, variations in the cloud top altitude along with the local temperature at the cloud top could result in variations in the continuum flux and might also allow some variation in the FeH band depth if clouds are sufficiently thin. No conclusive variability has been observed for these mid-T dwarfs, although there were small trends observed for the T4.5 2MASS J05591914-1404488 \citep{buenzli14, radigan14a}, which has a significant FeH feature. We note that a heterogeneous cloud distribution does not necessarily have to result in variability if the distribution is rotationally symmetric, distributed in many small patches, or if the object is seen close to pole-on. 

At the onset of the L/T transition, a spatially heterogeneous decrease in $f_{sed}$ between Luhman~16A and B appears to be responsible for the spectral changes and the observed variability, but the remaining clouds are still too thick to affect the FeH strength. It may be reasonable to assume that a similar combination of both mechanisms happens toward the end of the transition as well. Only more sophisticated modeling accounting for the columns of gaseous FeH above the cloud base and variations in the cloud top altitudes could rigorously test these conceptual models for the re-emergence of the FeH feature at mid-T type. Such models must account for possible heterogeneities in the different cloud layers, perhaps with varying horizontal distributions. Such a detailed study is well beyond the scope of this paper but provides fruitful possibilities for future research. 

For Luhman~16A and B, the wavelength dependence of the variability amplitude, i.e. the different amplitude in the water band but otherwise overall relatively gray variability, strongly points toward cloud top variations in the silicate cloud layer. Models with cloud holes, i.e. a mixture of a cloudy and clear photosphere, clearly cannot explain the observed variability, while a mixture of thinner and thicker clouds, i.e. cloud top variations, can.  In the 1.4 $\mu$m water band the flux emerges from well above the silicate cloud top. The reduced variability in that water band is therefore fully consistent with the thin/thick model. The model also predicts the same light curve phase at all wavelengths except in the deep K I line at 0.77 $\mu$m, while anti-correlation in the water band with respect to the J band peak might occur for some combinations of cloudy/clear models. 

A $\sim7$\% flux variation at  $1\,\rm\mu m$ can be produced by a cloud top temperature difference of about 14~K, just by the change in the Planck function. Given the local thermal profile, this corresponds to only a $\sim$~0.1 bar variation in the cloud top pressure. Such a scenario is also broadly consistent with the observed differences in the variability amplitude in the very deep K I line, which is extremely sensitive to pressure. 

Finally, there is also the possibility that the observed variability could stem solely from thermal variations \citep{robinson14} rather than cloud height variations. However, in this case there is not an obvious mechanism to maintain such a temperature anomaly, and mixed cloud properties are needed to reproduce the average spectrum of Luhman~16B. Therefore, our new observations further support the theory of silicate cloud height variations as the dominant reason for the variability we observe between 0.8 and 1.7 $\mu$m. 

\section{Conclusions}

We have obtained the first observations of spectroscopic variability of any brown dwarf in the 0.8-1.15~$\mu$m range, covering the FeH feature, by observing the L/T transition binary Luhman 16 with HST/WFC3. Both brown dwarfs display significant variability with likely periods of about 5~h, even Luhman~16A, which was previously found non-variable in the near-IR and in a partially overlapping wavelength region \citep{buenzli15}, and tentatively variable at i and z' bands \citep{biller13}. With both objects variable at least some of the time, the interpretation of spatially unresolved observations is more complicated than previously thought, and Luhman~16A cannot necessarily be used as a non-variable reference in spatially resolved observations. Finding two high-amplitude variables in a binary is somewhat unusual, as the independent probabilities for a late-L and early-T to be variable are only on the order of about 10\%. The Luhman 16 system may therefore have a common property that favors variability, although it is yet unclear what that common factor is. 

We show for the first time that the wavelength dependence of the variability amplitude from 0.8 to 1.15 $\mu$m is largely flat. This is the case for both objects. It can be explained by cloud height variations in the silicate clouds, while we can exclude holes. A similar conclusion was already reached for two other early T dwarfs from color variations \citep{apai13} and for Luhman~16B from the 1.1-1.65~$\mu$m observations \citep{buenzli15}, but here we can for the first time fit the 0.8-1.65~$\mu$m wavelength range covering many atomic and molecular features and obtain a consistent result. It suggests that the spectral changes across the L/T transition, at least from late-L to early-T, do not stem from the opening of deep holes, but from cloud thinning that can be spatially heterogeneous. 

We have also re-evaluated the role of FeH as a tracer of cloud holes, as first suggested by \citet{burgasser02}, or as a tracer of changing $f_{\mathrm{sed}}$ value as suggested by \citet{cushing08}. The models using the most recent opacities suggest that even models with a clear photosphere at 1200~K reach only the top of the Fe cloud at 1~$\mu$m and observations therefore cannot constrain whether there are holes in the Fe cloud. However, this also suggests that the re-emergence of the FeH feature through the L/T transition does not stem from holes in the Fe cloud. The reason for the re-emergence of FeH, in particular for the mid-T dwarfs, is therefore still not fully solved. Vertical mixing was already excluded by \citet{burgasser02} and \citet{lodders06}. It is still possible that a decrease in cloud opacity or variations in the cloud height and temperature in the silicate and/or iron cloud might affect the depth of the FeH feature for mid-Ts due to more of the FeH gas above the Fe cloud becoming visible. Detailed model calculations are required that study the effect of the different cloud layers and include a detailed FeH abundance profile between the clouds. However, for our late-L and early-T dwarf, where the cloud opacity is still very significant, the variability is dominated by cloud-height variations in the silicate cloud, and these are not sufficiently large to also affect the FeH strength.  The fact that Luhman 16 A and B have approximately the same FeH strength \citep{faherty14}, do not show any variability in FeH, but certainly have patchy clouds of different thickness, clearly indicates that the changing FeH strength does not correlate with other cloud evolution tracers at the early L/T transition.

Nevertheless, our observations have provided very valuable input into the structure of the silicate clouds at the transition from L to T-type brown dwarf. With their unique brightness, Luhman 16 A and B continue to be the prime target for detailed atmospheric studies of L/T transition brown dwarfs. The fact that both objects show variability at least some of the time provides a window into the three-dimensional atmospheric structure for two co-eval objects. The James Webb Space Telescope (JWST) will be able to precisely measure their variability over a broader wavelength range and at higher spectral resolution. Although the projected separation of the binary is currently decreasing and it may soon not be possible to spatially resolve them, it is expected that the projected separation will be increasing again by the time JWST launches and will be large enough to spatially resolve the binary. A potential orbit suggested by \citet{sahlmann15} suggests closest approach in 2017 and a separation of $>0.5''$ by 2018, increasing rapidly thereafter. 

Furthermore, astrometric monitoring will yield dynamical masses that can be used as independent inputs into atmospheric and evolutionary models. \citet{sahlmann15} already provide a first estimate of the system mass ratio of $q = 0.78\pm0.10$, confirming the somewhat lower mass of the further evolved B companion. However, the luminosities of the two brown dwarfs appear to be nearly equal \citep{faherty14, lodieu15}. The system therefore likely has a very shallow mass-luminosity relation, similar to the L/T transition binary SDSS~J105213.51+442255.7 \citep{dupuy15}. This lends further support to the necessity for evolutionary models to take into account cloud evolution at the L/T transition, as done for the hybrid tracks of \citet{saumon08}. Individual masses will be determined once about half of the binary orbit is covered. The orbit is likely highly inclined and a full orbital period may be about 45 years \citep{sahlmann15}. Although it may take about 20 years, the Luhman 16AB system will certainly become an extremely valuable benchmark L/T transition binary. Together with the variability, which we have now shown to be present for both objects and across a broad wavelength range, it will allow for detailed testing of atmospheric and evolutionary models that take into account the heterogeneous 3D atmospheric structure and evolution. 

\acknowledgements
We thank the anonymous referee for a helpful report. We  thank  the  staff  at the Space  Telescope  Science  Institute (STScI), in particular Amber Armstrong, for the scheduling of the observations. We thank Adam Burgasser, Alexei Kniazev, and Jackie Faherty for providing their published spectra of Luhman 16AB in electronic form. EB was supported by the Swiss National Science Foundation (SNSF). Based on observations made with the NASA/ESA Hubble Space Telescope, obtained at the Space Telescope Science Institute, which is operated by the Association of Universities for Research in Astronomy, Inc., under NASA contract NAS 5-26555. These observations are associated with program 13640. Support for program 13640 was provided by NASA through a grant from the Space Telescope Science  Institute. This research has made use of the SIMBAD database, operated at CDS, Strasbourg, France, and of NASA's Astrophysics Data System Bibliographic Services.

\end{document}